\documentclass[a4paper,aps,prd,10pt,preprintnumbers,showpacs,twocolumn,superscriptaddress,nofootinbib]{revtex4-1}
\usepackage{orcidlink,graphicx,cmap,amsmath}
\usepackage[utf8]{inputenc}
\usepackage[T1]{fontenc}

\def\re#1{Re(#1)}

\def\Order#1{{\cal O}\left(#1\right)}
\def\tr{\widetilde{\rho}}

\DeclareMathOperator\HF{{}_2F_1}

\begin{document}

\title{Charged black hole surrounded by a galactic halo in a de Sitter universe}

\author{R. A. Konoplya \orcidlink{0000-0003-1343-9584}}
\email{roman.konoplya@gmail.com}
\affiliation{Research Centre for Theoretical Physics and Astrophysics, Institute of Physics, Silesian University in Opava,\\ Bezručovo náměstí 13, CZ-74601 Opava, Czech Republic}

\author{Z. Stuchlík \orcidlink{0000-0003-2178-3588}}
\email{zdenek.stuchlik@physics.slu.cz}
\affiliation{Research Centre for Theoretical Physics and Astrophysics, Institute of Physics, Silesian University in Opava,\\ Bezručovo náměstí 13, CZ-74601 Opava, Czech Republic}

\author{A. Zhidenko \orcidlink{0000-0001-6838-3309}}
\email{olexandr.zhydenko@ufabc.edu.br}
\affiliation{Research Centre for Theoretical Physics and Astrophysics, Institute of Physics, Silesian University in Opava,\\ Bezručovo náměstí 13, CZ-74601 Opava, Czech Republic}
\affiliation{Centro de Matemática, Computação e Cognição (CMCC), Universidade Federal do ABC (UFABC),\\ Rua Abolição, CEP: 09210-180, Santo André, SP, Brazil}

\begin{abstract}
Assuming a sufficiently general form for the matter distribution function of a galactic halo, we have derived solutions to the Einstein-Maxwell equations describing a charged black hole embedded in such a halo, while also allowing for a nonzero cosmological constant. These solutions generalize our earlier results for neutral black holes in asymptotically flat spacetime. As specific realizations of the general distribution, we consider the Hernquist, Navarro-Frenk-White, Burkert, Taylor-Silk, and Moore halo profiles, thereby capturing a broad range of astrophysically motivated scenarios.
\end{abstract}

\pacs{04.70.Bw,95.35.+d,98.62.Js}

\maketitle

\section{Introduction}

Observational evidence strongly suggests that nearly every large galaxy hosts a supermassive black hole at its center~\cite{Kormendy:2013dxa}. These central black holes are embedded within complex astrophysical environments dominated by dark matter halos and, in many cases, shaped by large-scale cosmological dynamics. Consequently, the dynamics of matter, processes of radiation and various optical phenomena are broadly studied in the context of black holes immersed in the dark matter halo \cite{Dubinsky:2025fwv,Konoplya:2021ube,Liu:2024xcd,Macedo:2024qky,Heydari-Fard:2024wgu,Tan:2024hzw,Chen:2024lpd,Liu:2024bfj,Mollicone:2024lxy,Pezzella:2024tkf,Hamil:2025pte,Liu:2023vno,Datta:2023zmd,Zhao:2023itk,Chen:2023akf,Zhao:2023tyo,Xavier:2023exm,Figueiredo:2023gas,Daghigh:2022pcr,Liu:2022ygf,Liu:2022lrg,Jusufi:2022jxu,Igata:2022rcm,Stuchlik:2021gwg,Zhang:2021bdr,Konoplya:2019sns,Jha:2025uie,Santos:2025sun,Ahmed:2025ttq,Luo:2025xjb,Jha:2025xjf,Haroon:2025rzx,Konoplya:2025mvj,He:2025rjq,He:2024amh,Zare:2024dtf,Becar:2023zbl,Pantig:2022sjb,Stuchlik:2022xtq,Ravanal:2024odh,Pantig:2024rmr,Ovgun:2025bol}.

In order to model such systems more realistically, it is essential to account for both the influence of extended galactic matter distributions and the presence of a cosmological constant, which introduces a de Sitter-like asymptotic structure to the spacetime.

Astrophysical black holes are expected to have near-zero net electric charge, since any significant charge is rapidly neutralized by the surrounding plasma through selective accretion of opposite charges \cite{Bekenstein:1971ej}. However, a black hole immersed in a magnetized plasma can acquire a non-negligible induced charge due to charge separation effects, beyond the minimal equilibrium value \cite{Wald:1974np}. This enhanced charge can influence plasma motion near the black hole \cite{Ruffini:1975ne} and may help explain hot-spot flare dynamics observed near \textit{Sgr A*} \cite{Tursunov:2019mox}. Estimates of the astrophysically admissible electric charge of black holes can be found in \cite{Juraev:2024dju}.

In this work, we construct and analyze a family of spherically symmetric solutions to the Einstein-Maxwell equations representing charged black holes surrounded by a galactic halo and embedded in a de Sitter universe. By assuming a sufficiently general class of halo density profiles that includes the well-known Hernquist~\cite{Hernquist:1990be}, Burkert~\cite{Burkert:1995yz}, Navarro-Frenk-White~\cite{Navarro:1994hi}, Taylor-Silk~\cite{Taylor:2002zd}, and Moore et~al.~\cite{Moore:1997sg} models as particular cases, we obtain a class of analytic solutions that generalize our earlier results for neutral black holes in asymptotically flat spacetimes~\cite{Konoplya:2022hbl}.

The halo is modeled as an anisotropic fluid with purely tangential pressure and a density distribution that vanishes smoothly at both the event and cosmological horizons, ensuring regularity of the metric functions. In addition to incorporating electric charge and a cosmological constant, our solutions allow for a wide range of physically motivated halo configurations. We analyze the properties of these spacetimes and compute observable characteristics such as the black hole shadow, which may serve as potential discriminators between different halo models in future high-resolution observations.

The paper is organized as follows: in Sec.~\ref{sec:solution} we present the construction of the solutions for the charged black hole surrounded by galactic matter in the presence of a cosmological constant and derive the corresponding metric functions for various halo models. Sec.~\ref{sec:models} is devoted to a discussion of features of the solutions for various particular models of halo. In Sec.~\ref{sec:shadows} we calculate the radii of shadows cast by these black holes. The concluding section outlines possible extensions of our work.

\section{The matter distribution and solutions to the Einstein-Maxwell equations}\label{sec:solution}

Galactic matter is usually modeled by an anisotropic fluid with some density distribution, which implies an almost spherical halo dominated by dark matter \cite{Benson:2010de}. Depending on the size, mass, and form of a galaxy one or another distribution is preferable.
The generic density distribution of a galactic halo has the following form \cite{Dehnen:1993uh,Taylor:2002zd}:
\begin{equation}\label{density}
\rho(r)=2^{(\gamma-\alpha)/k}\rho_a \left(\frac{r}{a}\right)^{-\alpha} \left(1+\frac{r^k}{a^k}\right)^{-(\gamma-\alpha)/k}.
\end{equation}
The constant $\rho_a\equiv\rho(a)$ fixes the total mass of the galaxy,
\begin{equation}\label{totalhalomass}
M=4\pi\intop_0^s\rho(r)r^2dr,
\end{equation}
and $s$ is the radius of the halo, such that $s>a.$ Here. $a$ is a characteristic scale of the galactic halo. We also assume that $\alpha<3$ so that $M$ is finite.

In order to simplify notations we employ the general form for the line element,
\begin{eqnarray}\label{line-element}
ds^2&\equiv& g_{\mu\nu}dx^{\mu}dx^{\nu}
\\\nonumber
&=&-f(r)dt^2+\frac{B^2(r)}{f(r)}dr^2+r^2(d\theta^2+\sin^2\theta d\varphi^2),
\end{eqnarray}
where we assume that $B(r)>0$.

It is convenient to introduce the mass function, \mbox{$m(r)\leq r/2$}, such that
\begin{equation}\label{massfunction}
1-\frac{2m(r)}{r}\equiv\frac{f(r)}{B^2(r)}.
\end{equation}

We also consider the spherically symmetric electric field, which is determined by the following 4-potential:
\begin{equation}\label{Maxwellpotential}
A_{\mu}dx^{\mu}=V(r)dt.
\end{equation}

The galactic halo can be considered electrically neutral on large scales. Its dominant component---dark matter---is noninteracting and electrically neutral by definition, while the baryonic gas, though ionized, remains quasineutral due to rapid local charge compensation. Observationally, there is no evidence for large-scale electric fields in galactic halos, supporting the assumption of neutrality~\cite{Bertone:2004pz,Tumlinson_2017}.

Therefore, assuming that the environment is electrically neutral, the Maxwell equations,
\begin{equation}\label{Maxwell}
\frac{1}{\sqrt{-g}} \partial_\nu \left( \sqrt{-g} \, F^{\mu\nu} \right) = 0, \quad F_{\mu\nu} \equiv \partial_\mu A_\nu - \partial_\nu A_\mu,
\end{equation}
are reduced to the following form:
\begin{equation}\label{Veq}
V''(r)=\left(\frac{B'(r)}{B(r)}-\frac{2}{r}\right)V'(r).
\end{equation}
Thus, the electric field strength satisfies
\begin{equation}\label{Vex}
V'(r)=-\frac{Q}{r^2}B(r),
\end{equation}
where $Q$ is the integration constant. Since asymptotically $B(r)\to1$, we conclude that $Q$ is the electric charge.

Finally, we assume that the line element (\ref{line-element}) must be the solution to the Einstein equations with
\begin{itemize}
\item anisotropic matter of density $\rho(r)$,
\item only the tangential pressure $P(r)$,
\item the Maxwell field (\ref{Vex}),
\item cosmological constant $\Lambda$.
\end{itemize}

Then, the nonzero components of the corresponding stress-energy tensor are
\begin{equation}\label{stress-energy}
\begin{array}{rcl}
 T_{t}^{t} &=&-8\pi\rho(r)-\dfrac{Q^2}{r^4},\\
 T_{r}^{r} &=& -\dfrac{Q^2}{r^4}, \\
 T_{\theta}^{\theta} = T_{\varphi}^{\varphi} &=& 8\pi P(r)+\dfrac{Q^2}{r^4}.
\end{array}
\end{equation}

After some algebra, the Einstein equations are reduced to the following forms:
\begin{eqnarray}
\label{mder}
m'(r)&=&4\pi r^2\rho(r)+\frac{Q^2}{2r^2}+\frac{\Lambda r^2}{2},\\
\label{Bder}
B'(r)&=&4\pi r^2\rho(r)\frac{B(r)}{r-2m(r)},\\
\label{Pex}
P(r)&=&\frac{2m(r)r-Q^2+\Lambda r^4}{r-2m(r)}\frac{\rho(r)}{4r}.
\end{eqnarray}

It is important to note that the positive-definite density function (\ref{density}) is incompatible with the presence of the horizons. Indeed, from Eq.~(\ref{massfunction}) it follows that at the event horizon $r_0$, the following condition must be satisfied:
\begin{equation}\label{eventhorizon}
2m(r_0)=r_0.
\end{equation}
Then, since we assume that
$B(r_0)>0$, and Eq.~(\ref{density}) implies that $\rho(r_0)>0$, from Eq.~(\ref{Bder}) it follows that
\[\lim_{r\to r_0}B'(r)=\infty.\]

Thus, from (\ref{Bder}) it follows that $\rho(r)$ must vanish at the horizon, because, otherwise, $B(r)$ diverges. We also notice that the positive $\Lambda$-term leads to the asymptotically de Sitter solution, because, for large $r$, the mass function grows as
\[m(r)=\frac{\Lambda r^3}{6}+\Order{1},\]
which is faster than linear, so that there is a cosmological horizon, corresponding to a real $r_c>r_0$ such that
\begin{equation}\label{cosmologicalhorizon}
2m(r_c)=r_c.
\end{equation}
Again, it follows from (\ref{Bder}) that $B'(r)$ diverges as $r\to r_c$.

Thus, in order to obtain a black hole solution in a de Sitter universe with a spherical halo, we must modify the density $\rho\to\tr$ in such a way that it vanishes at both horizons. Following \cite{Konoplya:2022hbl}, we assume that there is no halo for $r>s$ and introduce a prefactor, which ensures that the density and its first $n$ derivatives vanish at the horizon. The latter conditions can be satisfied by the following density function:
\begin{equation}\label{modifieddensity}
  \tr(r)=\left\{\begin{array}{ll}
    \left(1-\dfrac{r_0}{r}\right)^{n+1}\rho(r), & r\leq s; \\
    0, & r>s.
  \end{array}\right.
\end{equation}

Now we are in the position to solve the Einstein equations. First, we solve Eq.~(\ref{mder}) with the initial condition (\ref{eventhorizon}), finding, thereby, the mass function
\begin{widetext}
\begin{equation}\label{msol}
  m(r)=\left\{
  \begin{array}{ll}
    \dfrac{r_0}{2}+\dfrac{Q^2}{2r_0}-\dfrac{Q^2}{2r}+\Lambda\dfrac{r^3-r_0^3}{6}+\sigma F\left(\dfrac{r}{a}\right)-\sigma F\left(\dfrac{r_0}{a}\right), &\qquad r<s;\\
    \dfrac{r_0}{2}+\dfrac{Q^2}{2r_0}-\dfrac{Q^2}{2r}+\Lambda\dfrac{r^3-r_0^3}{6}+\sigma F\left(\dfrac{s}{a}\right)-\sigma F\left(\dfrac{r_0}{a}\right), &\qquad r\geq s.
  \end{array}
\right.
\end{equation}
Here, we introduced
\begin{equation}\label{sigma}
\sigma\equiv4\pi\rho_a a^3, \qquad F\left(\dfrac{r}{a}\right)\equiv\dfrac{1}{a^3}\int r^2\tr(r)dr,
\end{equation}
so that $F(z)$ depends on the dimensionless parameter $r_0/a$ and on the halo parameters $\alpha$, $\gamma$, and $k$. We note that
\begin{equation}
F(z)=2^{\frac{\gamma-\alpha}{k}}\frac{z^{3-\alpha}}{3-\alpha}\HF\left(\dfrac{3-\alpha}{k},\dfrac{\gamma-\alpha}{k},\dfrac{3-\alpha}{k}+1;-z^k\right)+\Order{\frac{r_0}{a}}.
\label{Fdef}
\end{equation}
\end{widetext}
Here we follow the standard notation for the hypergeometric function
\[\HF\left(a,b,c;x\right)\equiv\sum_n\frac{\Gamma(a+n)\Gamma(b+n)\Gamma(c)}{\Gamma(a)\Gamma(b)\Gamma(c+n)}\cdot\frac{x^n}{n!}.\]

Similarly to the asymptotically flat case \cite{Konoplya:2022hbl}, once we assume that $r_0\ll a$ and neglect all the terms of the order $\Order{r_0/a}$, the solution (\ref{msol}) does not depend on the value of $n$ in (\ref{modifieddensity}).

The last two terms in the mass function~(\ref{msol}) correspond to the mass of the galactic halo,
\begin{eqnarray}
M&=&\sigma F\left(\dfrac{s}{a}\right)-\sigma F\left(\dfrac{r_0}{a}\right)
\\\nonumber &=&4\pi\intop_{r_0}^s r^2\tr(r)dr=4\pi\intop_{0}^s r^2\rho(r)dr+\Order{\frac{r_0}{a}},
\end{eqnarray}
which coincides with that given by Eq.~(\ref{totalhalomass}) if one neglects the terms of the order $\Order{r_0/a}$.

Using Eq.~(\ref{cosmologicalhorizon}), we relate the value of the cosmological constant $\Lambda$ and the cosmological horizon $r_c$,
\begin{equation}\label{cosmological}
\Lambda=\frac{6}{r_c^3-r_0^3}\left(\frac{r_c-r_0}{2}+\frac{Q^2}{2r_c}-\frac{Q^2}{2r_0}-M\right).
\end{equation}
In the limit of vanishing mass of the halo, $M=0$, Eq.~(\ref{cosmological}) is reduced to the relation between $\Lambda$ and $r_c$ for the Reissner-Nordström-de Sitter spacetime.

Finally, we solve Eq.~(\ref{Bder}) with the initial condition,
\begin{equation}\label{Binit}
B(r_c)=1,
\end{equation}
which corresponds to the Reissner-Nordström-de Sitter geometry at the cosmological horizon,
\begin{equation}\label{Bsol}
B(r)=\exp\left(-\intop_{r}^{r_c}\frac{4\pi x^2\tr(x)dx}{x-2m(x)}\right).
\end{equation}
The latter solution for $B(r)$ leads to the following expression for the lapse function:
\begin{equation}\label{fsol}
f(r)=\exp\left(-\intop_{r}^{r_c}\frac{8\pi x^2\tr(x)dx}{x-2m(x)}\right)\left(1-\frac{2m(r)}{r}\right).
\end{equation}

We also choose the standard Reissner-Nordström-de Sitter gauge for the electric potential (\ref{Vex}), $V(r_c)=Q/r_c$,
\begin{equation}
V(r)=\intop_{r}^{r_c}\frac{Q B(x)dx}{x^2}+\frac{Q}{r_c}.
\end{equation}

We assume that the size of the halo is smaller than the cosmological radius, $s<r_c$. It follows from the above conditions that, for $s<r<r_c$, we have Reissner-Nordström-de Sitter geometry with $B(r)=1$ and $V(r)=Q/r$.

\section{Particular models of galactic halo}\label{sec:models}

Here we will write out the explicit expressions for the mass of the halo and the function $F(z)$ for the particular models of the density profiles for the galactic halo: Hernquist~\cite{Hernquist:1990be}, Burkert~\cite{Burkert:1995yz}, Navarro-Frenk-White~\cite{Navarro:1994hi}, Taylor-Silk~\cite{Taylor:2002zd}, and Moore et~al.~\cite{Moore:1997sg} models.

\begin{figure*}
\resizebox{\linewidth}{!}{\includegraphics{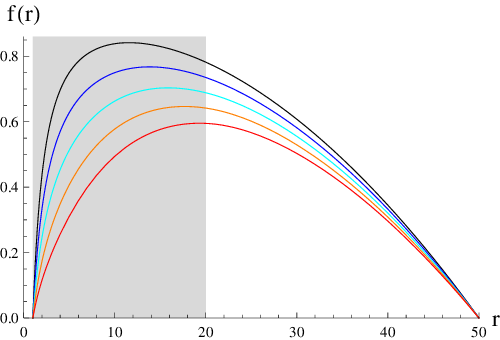}\includegraphics{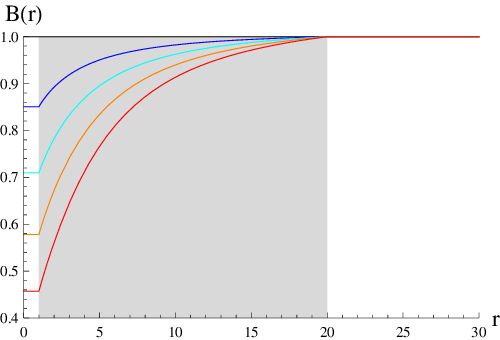}}
\caption{Metric functions for the charged black hole in de Sitter universe ($Q=0.5$, $r_0=1$, $r_c=50$) with the Hernquist halo ($\alpha=1$, $\gamma=4$, $k=1$, $n=0$) for $a=5$ and $s=20$ of various masses: (from top to bottom) $M=0$ (black, Reissner-Nordström-de Sitter solution), $M=0.5$ (blue), $M=1$ (cyan), $M=1.5$ (orange), and $M=2$ (red). Shaded region corresponds to the nonzero density of the dark matter (halo).}\label{fig:Hernquistmetric}
\end{figure*}

\begin{itemize}
\item \emph{Hernquist model}: $\alpha=1$, $\gamma=4$, $k=1$,
\begin{eqnarray}
\nonumber F(z)&=&\frac{4z^2}{(1+z)^2},\\
\nonumber M&=&\frac{4\sigma s^2}{a^2+s^2}-\frac{4\sigma r_0^2}{a^2+r_0^2}.
\end{eqnarray}
For $r<s$ we obtain
\begin{eqnarray}\nonumber
1-\frac{2m(r)}{r}&=&\left(1-\frac{r_0}{r}\right)\Biggl(1-\frac{Q^2}{rr_0}-\frac{\Lambda(r^2+rr_0+r_0^2)}{3}
\\\nonumber
&&-\frac{8\sigma a^2r}{(a+r)^2(a+r_0)^2}\left(1+\frac{2r_0}{a}+\frac{r_0}{r}\right)\Biggr),
\end{eqnarray}
coinciding with Eq.~(33) of~\cite{Konoplya:2022hbl} when we neglect terms of the order $\Order{r_0/a}$ and take $\Lambda=0$ and $Q=0$.

\item \emph{Navarro-Frenk-White model}: $\alpha=1$, $\gamma=3$, $k=1$,
\begin{eqnarray}
\nonumber F(z)&=&4\left(\ln(1+z)-z\right),\\
\nonumber M&=&4\sigma\left(\ln\left(\frac{a+s}{a+r_0}\right)-\frac{a(s-r_0)}{(a+r_0)(a+s)}\right).
\end{eqnarray}
For $r<s$ the mass function takes the form
\begin{eqnarray}\nonumber
1-\frac{2m(r)}{r}&=&\left(1-\frac{r_0}{r}\right)\Biggl(1-\frac{Q^2}{rr_0}-\frac{\Lambda(r^2+rr_0+r_0^2)}{3}
\\\nonumber
&&-\frac{8\sigma a}{(a+r)(a+r_0)}\Biggr)+\frac{8\sigma}{r}\ln\left(\frac{a+r_0}{a+r}\right).
\end{eqnarray}
Again, neglecting terms of the order $\Order{r_0/a}$ and taking $\Lambda=0$ and $Q=0$, we reproduce Eq.~(3) of~\cite{Konoplya:2022hbl}.

\item \emph{Burkert model}: $\alpha=1$, $\gamma=3$, $k=2$,
\begin{eqnarray}
\nonumber F(z)&=&\ln(1+z^2),\\
\nonumber M&=&\sigma\ln\left(\frac{a^2+s^2}{a^2+r_0^2}\right).
\end{eqnarray}
For $r<s$ we have
\begin{eqnarray}\nonumber
1-\frac{2m(r)}{r}&=&\left(1-\frac{r_0}{r}\right)\left(1-\frac{Q^2}{rr_0}-\frac{\Lambda(r^2+rr_0+r_0^2)}{3}\right)
\\\nonumber
&&+\frac{2\sigma}{r}\ln\left(\frac{a^2+r_0^2}{a^2+r^2}\right),
\end{eqnarray}
which is reduced to Eq.~(4) of~\cite{Konoplya:2022hbl} after neglecting terms of the order $\Order{r_0/a}$ and taking $\Lambda=0$ and $Q=0$.

\item \emph{Taylor-Silk model}: $\alpha=3/2$, $\gamma=3$, $k=3/2$,
\begin{eqnarray}
\nonumber F(z)&=&\frac{4}{3}\ln(1+z^{3/2}),\\
\nonumber M&=&\frac{4\sigma}{3}\ln\left(\frac{a^{3/2}+s^{3/2}}{a^{3/2}+r_0^{3/2}}\right).
\end{eqnarray}
For $r<s$ the mass function takes the following form:
\begin{eqnarray}\nonumber
1-\frac{2m(r)}{r}&=&\left(1-\frac{r_0}{r}\right)\left(1-\frac{Q^2}{rr_0}-\frac{\Lambda(r^2+rr_0+r_0^2)}{3}\right)
\\\nonumber
&&+\frac{8\sigma}{3r}\ln\left(\frac{a^{3/2}+r_0^{3/2}}{a^{3/2}+r^{3/2}}\right).
\end{eqnarray}

\item \emph{Moore model}: $\alpha=7/5$, $\gamma=14/5$, $k=7/5$.

In this case, we are unable to obtain the mass function and the total mass of the galaxy in terms of the elementary functions, however, equation~(\ref{msol}) provides the expressions in terms of the hypergeometric functions.
\end{itemize}

Unlike for the asymptotically flat case one cannot take the limit $s\to\infty$ even for the $\gamma>3$ (e.~g., for the Hernquist profile). Although the mass of the galaxy remains finite the integral (\ref{Bsol}) diverges. Here, we are restricted by the additional condition $s<r_c$.

In Fig.~\ref{fig:Hernquistmetric} we show the behavior of the metric functions for the black holes with the Hernquist halo.
In the ancillary \emph{Mathematica\textregistered{}} Notebook, we share the numerical solutions for arbitrary density profiles for the galactic halos, including (but not limited to) all the models considered above.\footnote{The ancillary \emph{Mathematica\textregistered{}} Notebook file is available from \url{https://arxiv.org/src/2509.03301v1/anc}.}

\section{Shadow radius}\label{sec:shadows}

Here we will compare the obtained metrics with the analytic approximations in the corresponding asymptotically flat spacetimes \cite{Konoplya:2022hbl}. We calculate a simple observable characteristic, namely the shadow radius $R_s$, which corresponds to the minimum of the following function:
\begin{equation}\label{photon}
R_{s}=\min_{r_0<r<r_c}\frac{r}{\sqrt{f(r)}}=\frac{r_{ph}}{\sqrt{f(r_{ph})}},
\end{equation}
where $r_{ph}$ is the value of the radial coordinate of a massless particle on the circular orbit. It is important to note that, since an asymptotic observer in a de Sitter universe cannot see the black hole, the actual radius of the observed shadow depends on the observer's position relative to the black hole. Nevertheless, the value of $R_s$ is gauge independent, and we use it for illustrative purposes.

Additionally, the value of $R_s$ is inversely proportional to the real part of the quasinormal frequencies \cite{Konoplya:2011qq,Bolokhov:2025uxz} in the eikonal regime measured in units of the coordinate time,
\begin{equation}
\re{\omega}=\frac{1}{R_s}\left(\ell+\frac{1}{2}\right)+\Order{\frac{1}{\ell}}.
\end{equation}

However, the above relation is based on the correspondence between eikonal quasinormal modes and null geodesics~\cite{Cardoso:2008bp}, and should be interpreted with care, as this correspondence breaks down in several important cases~\cite{Bolokhov:2023dxq}. In particular, it is modified for asymptotically de Sitter black holes~\cite{Konoplya:2022gjp,Konoplya:2019hml,Konoplya:2020bxa,Konoplya:2017wot}, where, in addition to the Schwarzschild-like branch of quasinormal modes, there exists a distinct de Sitter branch~\cite{Konoplya:2022xid,Konoplya:2024ptj} that does not obey the geodesic correspondence. In the relation above, the symbol $\omega$ refers specifically to the least damped frequency of the black hole branch of modes only (see Sec.~IV of~\cite{Stuchlik:2025mjj} for a discussion).

In \cite{Konoplya:2022hbl} the analytic approximation has been derived for the asymptotically flat spacetime
\begin{equation}\label{shadowapproximation}
\frac{R_S}{R_0}= 1+A_0\frac{M}{a}+\Order{\frac{r_0}{a}}+\Order{\frac{M}{a}}^2,
\end{equation}
where
\[R_0=\frac{3\sqrt{3}}{2}r_0\]
is the shadow radius of the Schwarzschild black hole and the coefficient $A_0$ can be approximated as
\begin{equation}\label{AA}
A_0\approx\dfrac{(3-\alpha)a}{(2-\alpha)s}\dfrac{\HF\left(\frac{2-\alpha}{k},\frac{\gamma-\alpha}{k},\frac{2-\alpha}{k}+1;-\left(\frac{s}{a}\right)^k\right)}{\HF\left(\frac{3-\alpha}{k},\frac{\gamma-\alpha}{k},\frac{3-\alpha}{k}+1;-\left(\frac{s}{a}\right)^k\right)}.
\end{equation}

\begin{figure}
\resizebox{\linewidth}{!}{\includegraphics{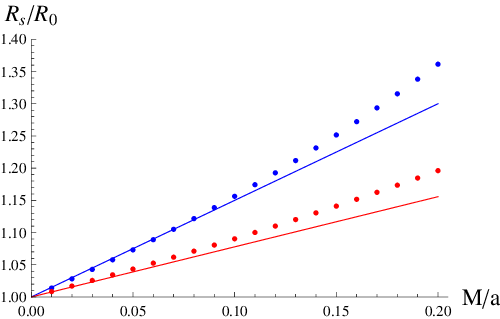}}
\caption{Shadow radius for the charged black hole in de Sitter universe ($Q=0.5$, $r_0=1$, $r_c=50$) with the Hernquist halo ($\alpha=1$, $\gamma=4$, $k=1$, $n=0$) for $a=5$ and $s=20$ (upper, blue) and with the Burkert halo ($\alpha=1$, $\gamma=3$, $k=2$, $n=0$) for $a=5$ and $s=30$ (lower, red). The shadow radius is measured in the units of the shadow radius for the corresponding Reissner-Nordström-de Sitter black hole \mbox{($R_0\approx2.8445$)}. Solid lines correspond to the analytic approximations \mbox{$R_s=R_0(1+A_0M/a)$}, where $A_0=1.5$ for the Hernquist halo and $A_0\approx0.78$ for the Burkert halo.}\label{fig:shadows}
\end{figure}

\begin{figure}
\resizebox{\linewidth}{!}{\includegraphics{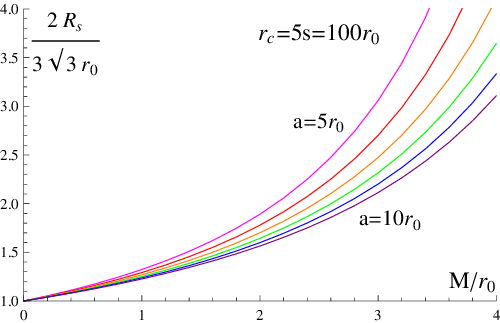}}
\caption{Shadow radius for the uncharged black hole in de Sitter universe ($Q=0$, $r_c=100r_0$) with the Hernquist model of the halo ($s=20r_0$). From top to bottom: $a=5r_0$, $a=6r_0$, $a=7r_0$, $a=8r_0$, $a=9r_0$, and $a=10r_0$. The shadow radius is measured in the units of the shadow of the Schwarzschild black hole of the same size.}\label{fig:shadows-Hernquist}
\end{figure}

It is important to emphasize that Eq.~(\ref{shadowapproximation}) was derived in \cite{Konoplya:2022hbl} for the case of an uncharged black hole in an asymptotically flat spacetime. Here, we assume that the leading correction to the shadow size induced by the galactic halo is again governed by the coefficient $A_0$ given in Eq.~(\ref{AA}). Therefore, one can estimate the corrected shadow radius $R_s$ by substituting, as $R_0$ in Eq.~(\ref{shadowapproximation}), the shadow radius of the corresponding Reissner-Nordström-de Sitter black hole. Such an approach is justified as long as the Reissner-Nordström-de Sitter geometry captures the primary effect of the black hole charge and cosmological constant, while the galactic halo is only a subleading correction. Moreover, as illustrated in Fig.~\ref{fig:shadows}, the analytic expression (\ref{shadowapproximation}) remains a reliable approximation, even when the galactic scale parameter $s$ is excessively large and comparable to the cosmological horizon radius $r_c$, confirming the robustness of the approximate formula of \cite{Konoplya:2022hbl}. Although it may appear unusual to consider a galactic extension comparable to the size of the cosmological horizon, we show this example for testing the validity of the approximation.

\begin{figure}
\resizebox{\linewidth}{!}{\includegraphics{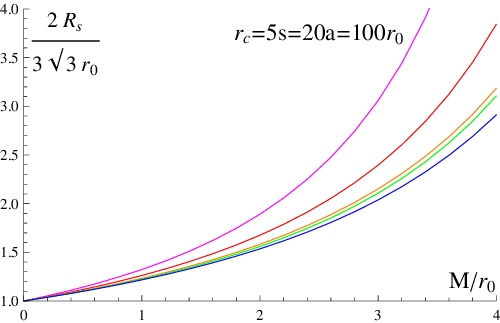}}
\caption{Shadow radius for the uncharged black hole in de Sitter universe ($Q=0$, $r_c=100r_0$) with various models of halo of the same size, $s=4a=20r_0$. From top to bottom:\\
{\textbullet} the Hernquist model with $\alpha=1$, $\gamma=4$, $k=1$, $n=0$ (magenta);\\
{\textbullet} the Taylor-Silk model with $\alpha=3/2$, $\gamma=3$, $k=3/2$, $n=0$ (red);\\
{\textbullet} the Moore model with $\alpha=7/5$, $\gamma=14/5$, $k=7/5$, $n=0$ (orange);\\
{\textbullet} the Burkert model with $\alpha=1$, $\gamma=3$, $k=2$, $n=0$ (green);\\
{\textbullet} the Navarro-Frenk-White model with $\alpha=1$, $\gamma=3$, $k=1$, $n=0$ (blue).
\\The shadow radius is measured in the units of the shadow of the Schwarzschild black hole of the same size.}\label{fig:shadows-a5}
\end{figure}

In general, the presence of a cosmological constant and a galactic halo increases the apparent radius of the shadow. When the halo mass is large, the shadow radius grows more rapidly. From Fig.~\ref{fig:shadows-Hernquist} we see that the shadow is also enlarged when the halo matter is concentrated closer to the black hole (i.e., when the parameter $a$ is small). For the asymptotically flat spacetime the same behavior was reported in \cite{Xavier:2023exm}. In Fig.~\ref{fig:shadows-a5}, we present the shadow radius, measured in units of the Schwarzschild shadow radius ($R_s=3\sqrt{3}r_0/2$), as a function of the galactic halo mass for various models. We observe that, qualitatively, the behavior is consistent across different models, with the shadow radius growing rapidly as the halo mass increases.

\section{Conclusions}

In this work, we have constructed and analyzed a new family of spherically symmetric black hole solutions representing a charged black hole immersed in a general galactic halo and embedded in a de Sitter universe. The galactic halo was modeled as an anisotropic fluid with purely tangential pressure and vanishing density at both the event and cosmological horizons, ensuring the regularity of the metric functions throughout the spacetime. Our approach accommodates a broad class of density profiles, including the Hernquist, Burkert, Taylor-Silk, and Moore models, as particular cases of a general distribution function. The analytic solutions for the mass function have been obtained for the above models.

The solutions we obtained generalize earlier results for neutral black holes in asymptotically flat spacetimes to the more general case of charged black holes in cosmological and astrophysical environments. We have also examined the influence of the halo parameters on the black hole shadow, demonstrating that the presence of halo matter and a nonzero cosmological constant leads to modifications in the shadow radius. These deviations may offer a way to constrain the properties of galactic halos and black hole charge using high-resolution observational data.

Our framework provides a consistent and physically motivated setting for studying black holes surrounded by galactic matter and opens the door to further extensions. In particular, it would be interesting to analyze the stability and quasinormal mode spectra of these solutions, as well as their gravitational lensing and accretion disk properties. The inclusion of rotation and more complex matter distributions also presents a natural avenue for future research.

\begin{acknowledgments}
A.~Z. acknowledges the Research Centre for Theoretical Physics and Astrophysics at the Institute of Physics, Silesian University in Opava, for their hospitality.
\end{acknowledgments}

\bibliography{RNdSHalo}

\begin{thebibliography}{69}%
\makeatletter
\providecommand \@ifxundefined [1]{%
 \@ifx{#1\undefined}
}%
\providecommand \@ifnum [1]{%
 \ifnum #1\expandafter \@firstoftwo
 \else \expandafter \@secondoftwo
 \fi
}%
\providecommand \@ifx [1]{%
 \ifx #1\expandafter \@firstoftwo
 \else \expandafter \@secondoftwo
 \fi
}%
\providecommand \natexlab [1]{#1}%
\providecommand \enquote  [1]{``#1''}%
\providecommand \bibnamefont  [1]{#1}%
\providecommand \bibfnamefont [1]{#1}%
\providecommand \citenamefont [1]{#1}%
\providecommand \href@noop [0]{\@secondoftwo}%
\providecommand \href [0]{\begingroup \@sanitize@url \@href}%
\providecommand \@href[1]{\@@startlink{#1}\@@href}%
\providecommand \@@href[1]{\endgroup#1\@@endlink}%
\providecommand \@sanitize@url [0]{\catcode `\\12\catcode `\$12\catcode
  `\&12\catcode `\#12\catcode `\^12\catcode `\_12\catcode `\%12\relax}%
\providecommand \@@startlink[1]{}%
\providecommand \@@endlink[0]{}%
\providecommand \url  [0]{\begingroup\@sanitize@url \@url }%
\providecommand \@url [1]{\endgroup\@href {#1}{\urlprefix }}%
\providecommand \urlprefix  [0]{URL }%
\providecommand \Eprint [0]{\href }%
\providecommand \doibase [0]{http://dx.doi.org/}%
\providecommand \selectlanguage [0]{\@gobble}%
\providecommand \bibinfo  [0]{\@secondoftwo}%
\providecommand \bibfield  [0]{\@secondoftwo}%
\providecommand \translation [1]{[#1]}%
\providecommand \BibitemOpen [0]{}%
\providecommand \bibitemStop [0]{}%
\providecommand \bibitemNoStop [0]{.\EOS\space}%
\providecommand \EOS [0]{\spacefactor3000\relax}%
\providecommand \BibitemShut  [1]{\csname bibitem#1\endcsname}%
\let\auto@bib@innerbib\@empty
\bibitem [{\citenamefont {Kormendy}\ and\ \citenamefont
  {Ho}(2013)}]{Kormendy:2013dxa}%
  \BibitemOpen
  \bibfield  {author} {\bibinfo {author} {\bibfnamefont {J.}~\bibnamefont
  {Kormendy}}\ and\ \bibinfo {author} {\bibfnamefont {L.~C.}\ \bibnamefont
  {Ho}},\ }\href {\doibase 10.1146/annurev-astro-082708-101811} {\bibfield
  {journal} {\bibinfo  {journal} {Ann. Rev. Astron. Astrophys.}\ }\textbf
  {\bibinfo {volume} {51}},\ \bibinfo {pages} {511} (\bibinfo {year} {2013})},\
  \Eprint {http://arxiv.org/abs/1304.7762} {arXiv:1304.7762 [astro-ph.CO]}
  \BibitemShut {NoStop}%
\bibitem [{\citenamefont {Dubinsky}(2025)}]{Dubinsky:2025fwv}%
  \BibitemOpen
  \bibfield  {author} {\bibinfo {author} {\bibfnamefont {A.}~\bibnamefont
  {Dubinsky}},\ }\href@noop {} {\bibfield  {journal} {\bibinfo  {journal}
  {International Journal of Gravitation and Theoretical Physics}\ }\textbf
  {\bibinfo {volume} {1}},\ \bibinfo {pages} {2} (\bibinfo {year} {2025})},\
  \Eprint {http://arxiv.org/abs/2507.00256} {arXiv:2507.00256 [gr-qc]}
  \BibitemShut {NoStop}%
\bibitem [{\citenamefont {Konoplya}(2021)}]{Konoplya:2021ube}%
  \BibitemOpen
  \bibfield  {author} {\bibinfo {author} {\bibfnamefont {R.~A.}\ \bibnamefont
  {Konoplya}},\ }\href {\doibase 10.1016/j.physletb.2021.136734} {\bibfield
  {journal} {\bibinfo  {journal} {Phys. Lett. B}\ }\textbf {\bibinfo {volume}
  {823}},\ \bibinfo {pages} {136734} (\bibinfo {year} {2021})},\ \Eprint
  {http://arxiv.org/abs/2109.01640} {arXiv:2109.01640 [gr-qc]} \BibitemShut
  {NoStop}%
\bibitem [{\citenamefont {Liu}\ \emph {et~al.}(2024{\natexlab{a}})\citenamefont
  {Liu}, \citenamefont {Yang},\ and\ \citenamefont {Long}}]{Liu:2024xcd}%
  \BibitemOpen
  \bibfield  {author} {\bibinfo {author} {\bibfnamefont {D.}~\bibnamefont
  {Liu}}, \bibinfo {author} {\bibfnamefont {Y.}~\bibnamefont {Yang}}, \ and\
  \bibinfo {author} {\bibfnamefont {Z.-W.}\ \bibnamefont {Long}},\ }\href
  {\doibase 10.1140/epjc/s10052-024-13096-8} {\bibfield  {journal} {\bibinfo
  {journal} {Eur. Phys. J. C}\ }\textbf {\bibinfo {volume} {84}},\ \bibinfo
  {pages} {731} (\bibinfo {year} {2024}{\natexlab{a}})},\ \Eprint
  {http://arxiv.org/abs/2401.09182} {arXiv:2401.09182 [gr-qc]} \BibitemShut
  {NoStop}%
\bibitem [{\citenamefont {Macedo}\ \emph {et~al.}(2024)\citenamefont {Macedo},
  \citenamefont {Rosa},\ and\ \citenamefont {Rubiera-Garcia}}]{Macedo:2024qky}%
  \BibitemOpen
  \bibfield  {author} {\bibinfo {author} {\bibfnamefont {C.~F.~B.}\
  \bibnamefont {Macedo}}, \bibinfo {author} {\bibfnamefont {J.~L.}\
  \bibnamefont {Rosa}}, \ and\ \bibinfo {author} {\bibfnamefont
  {D.}~\bibnamefont {Rubiera-Garcia}},\ }\href {\doibase
  10.1088/1475-7516/2024/07/046} {\bibfield  {journal} {\bibinfo  {journal}
  {JCAP}\ }\textbf {\bibinfo {volume} {07}},\ \bibinfo {pages} {046} (\bibinfo
  {year} {2024})},\ \Eprint {http://arxiv.org/abs/2402.13047} {arXiv:2402.13047
  [gr-qc]} \BibitemShut {NoStop}%
\bibitem [{\citenamefont {Heydari-Fard}\ \emph {et~al.}(2025)\citenamefont
  {Heydari-Fard}, \citenamefont {Heydari-Fard},\ and\ \citenamefont
  {Riazi}}]{Heydari-Fard:2024wgu}%
  \BibitemOpen
  \bibfield  {author} {\bibinfo {author} {\bibfnamefont {M.}~\bibnamefont
  {Heydari-Fard}}, \bibinfo {author} {\bibfnamefont {M.}~\bibnamefont
  {Heydari-Fard}}, \ and\ \bibinfo {author} {\bibfnamefont {N.}~\bibnamefont
  {Riazi}},\ }\href {\doibase 10.1007/s10714-025-03382-5} {\bibfield  {journal}
  {\bibinfo  {journal} {Gen. Rel. Grav.}\ }\textbf {\bibinfo {volume} {57}},\
  \bibinfo {pages} {49} (\bibinfo {year} {2025})},\ \Eprint
  {http://arxiv.org/abs/2408.16020} {arXiv:2408.16020 [gr-qc]} \BibitemShut
  {NoStop}%
\bibitem [{\citenamefont {Tan}\ \emph {et~al.}(2025)\citenamefont {Tan},
  \citenamefont {Deng}, \citenamefont {Long},\ and\ \citenamefont
  {Jing}}]{Tan:2024hzw}%
  \BibitemOpen
  \bibfield  {author} {\bibinfo {author} {\bibfnamefont {Q.}~\bibnamefont
  {Tan}}, \bibinfo {author} {\bibfnamefont {W.}~\bibnamefont {Deng}}, \bibinfo
  {author} {\bibfnamefont {S.}~\bibnamefont {Long}}, \ and\ \bibinfo {author}
  {\bibfnamefont {J.}~\bibnamefont {Jing}},\ }\href {\doibase
  10.1088/1475-7516/2025/05/044} {\bibfield  {journal} {\bibinfo  {journal}
  {JCAP}\ }\textbf {\bibinfo {volume} {05}},\ \bibinfo {pages} {044} (\bibinfo
  {year} {2025})},\ \Eprint {http://arxiv.org/abs/2409.17760} {arXiv:2409.17760
  [gr-qc]} \BibitemShut {NoStop}%
\bibitem [{\citenamefont {Chen}\ \emph {et~al.}(2024)\citenamefont {Chen},
  \citenamefont {Javed}, \citenamefont {Mustafa}, \citenamefont {Maurya},\ and\
  \citenamefont {Ray}}]{Chen:2024lpd}%
  \BibitemOpen
  \bibfield  {author} {\bibinfo {author} {\bibfnamefont {R.-Y.}\ \bibnamefont
  {Chen}}, \bibinfo {author} {\bibfnamefont {F.}~\bibnamefont {Javed}},
  \bibinfo {author} {\bibfnamefont {D.~G.}\ \bibnamefont {Mustafa}}, \bibinfo
  {author} {\bibfnamefont {S.~K.}\ \bibnamefont {Maurya}}, \ and\ \bibinfo
  {author} {\bibfnamefont {S.}~\bibnamefont {Ray}},\ }\href {\doibase
  10.1016/j.jheap.2024.09.010} {\bibfield  {journal} {\bibinfo  {journal}
  {JHEAp}\ }\textbf {\bibinfo {volume} {44}},\ \bibinfo {pages} {172} (\bibinfo
  {year} {2024})}\BibitemShut {NoStop}%
\bibitem [{\citenamefont {Liu}\ \emph {et~al.}(2025)\citenamefont {Liu},
  \citenamefont {Mu}, \citenamefont {Tao},\ and\ \citenamefont
  {Weng}}]{Liu:2024bfj}%
  \BibitemOpen
  \bibfield  {author} {\bibinfo {author} {\bibfnamefont {Y.}~\bibnamefont
  {Liu}}, \bibinfo {author} {\bibfnamefont {B.}~\bibnamefont {Mu}}, \bibinfo
  {author} {\bibfnamefont {J.}~\bibnamefont {Tao}}, \ and\ \bibinfo {author}
  {\bibfnamefont {Y.}~\bibnamefont {Weng}},\ }\href {\doibase
  10.1016/j.nuclphysb.2024.116787} {\bibfield  {journal} {\bibinfo  {journal}
  {Nucl. Phys. B}\ }\textbf {\bibinfo {volume} {1010}},\ \bibinfo {pages}
  {116787} (\bibinfo {year} {2025})},\ \Eprint
  {http://arxiv.org/abs/2409.20333} {arXiv:2409.20333 [gr-qc]} \BibitemShut
  {NoStop}%
\bibitem [{\citenamefont {Mollicone}\ and\ \citenamefont
  {Destounis}(2025)}]{Mollicone:2024lxy}%
  \BibitemOpen
  \bibfield  {author} {\bibinfo {author} {\bibfnamefont {A.}~\bibnamefont
  {Mollicone}}\ and\ \bibinfo {author} {\bibfnamefont {K.}~\bibnamefont
  {Destounis}},\ }\href {\doibase 10.1103/PhysRevD.111.024017} {\bibfield
  {journal} {\bibinfo  {journal} {Phys. Rev. D}\ }\textbf {\bibinfo {volume}
  {111}},\ \bibinfo {pages} {024017} (\bibinfo {year} {2025})},\ \Eprint
  {http://arxiv.org/abs/2410.11952} {arXiv:2410.11952 [gr-qc]} \BibitemShut
  {NoStop}%
\bibitem [{\citenamefont {Pezzella}\ \emph {et~al.}(2025)\citenamefont
  {Pezzella}, \citenamefont {Destounis}, \citenamefont {Maselli},\ and\
  \citenamefont {Cardoso}}]{Pezzella:2024tkf}%
  \BibitemOpen
  \bibfield  {author} {\bibinfo {author} {\bibfnamefont {L.}~\bibnamefont
  {Pezzella}}, \bibinfo {author} {\bibfnamefont {K.}~\bibnamefont {Destounis}},
  \bibinfo {author} {\bibfnamefont {A.}~\bibnamefont {Maselli}}, \ and\
  \bibinfo {author} {\bibfnamefont {V.}~\bibnamefont {Cardoso}},\ }\href
  {\doibase 10.1103/PhysRevD.111.064026} {\bibfield  {journal} {\bibinfo
  {journal} {Phys. Rev. D}\ }\textbf {\bibinfo {volume} {111}},\ \bibinfo
  {pages} {064026} (\bibinfo {year} {2025})},\ \Eprint
  {http://arxiv.org/abs/2412.18651} {arXiv:2412.18651 [gr-qc]} \BibitemShut
  {NoStop}%
\bibitem [{\citenamefont {Hamil}\ \emph {et~al.}(2025)\citenamefont {Hamil},
  \citenamefont {Al-Badawi},\ and\ \citenamefont
  {Lütfüoğlu}}]{Hamil:2025pte}%
  \BibitemOpen
  \bibfield  {author} {\bibinfo {author} {\bibfnamefont {B.}~\bibnamefont
  {Hamil}}, \bibinfo {author} {\bibfnamefont {A.}~\bibnamefont {Al-Badawi}}, \
  and\ \bibinfo {author} {\bibfnamefont {B.~C.}\ \bibnamefont {Lütfüoğlu}},\
  }\href@noop {} {\  (\bibinfo {year} {2025})},\ \Eprint
  {http://arxiv.org/abs/2505.18611} {arXiv:2505.18611 [gr-qc]} \BibitemShut
  {NoStop}%
\bibitem [{\citenamefont {Liu}\ \emph {et~al.}(2024{\natexlab{b}})\citenamefont
  {Liu}, \citenamefont {Yang},\ and\ \citenamefont {Long}}]{Liu:2023vno}%
  \BibitemOpen
  \bibfield  {author} {\bibinfo {author} {\bibfnamefont {D.}~\bibnamefont
  {Liu}}, \bibinfo {author} {\bibfnamefont {Y.}~\bibnamefont {Yang}}, \ and\
  \bibinfo {author} {\bibfnamefont {Z.-W.}\ \bibnamefont {Long}},\ }\href
  {\doibase 10.1140/epjc/s10052-024-13255-x} {\bibfield  {journal} {\bibinfo
  {journal} {Eur. Phys. J. C}\ }\textbf {\bibinfo {volume} {84}},\ \bibinfo
  {pages} {871} (\bibinfo {year} {2024}{\natexlab{b}})},\ \Eprint
  {http://arxiv.org/abs/2312.07074} {arXiv:2312.07074 [gr-qc]} \BibitemShut
  {NoStop}%
\bibitem [{\citenamefont {Datta}(2024)}]{Datta:2023zmd}%
  \BibitemOpen
  \bibfield  {author} {\bibinfo {author} {\bibfnamefont {S.}~\bibnamefont
  {Datta}},\ }\href {\doibase 10.1103/PhysRevD.109.104042} {\bibfield
  {journal} {\bibinfo  {journal} {Phys. Rev. D}\ }\textbf {\bibinfo {volume}
  {109}},\ \bibinfo {pages} {104042} (\bibinfo {year} {2024})},\ \Eprint
  {http://arxiv.org/abs/2312.01277} {arXiv:2312.01277 [gr-qc]} \BibitemShut
  {NoStop}%
\bibitem [{\citenamefont {Zhao}\ \emph {et~al.}(2024)\citenamefont {Zhao},
  \citenamefont {Sun}, \citenamefont {Cao}, \citenamefont {Lin},\ and\
  \citenamefont {Qian}}]{Zhao:2023itk}%
  \BibitemOpen
  \bibfield  {author} {\bibinfo {author} {\bibfnamefont {Y.}~\bibnamefont
  {Zhao}}, \bibinfo {author} {\bibfnamefont {B.}~\bibnamefont {Sun}}, \bibinfo
  {author} {\bibfnamefont {Z.}~\bibnamefont {Cao}}, \bibinfo {author}
  {\bibfnamefont {K.}~\bibnamefont {Lin}}, \ and\ \bibinfo {author}
  {\bibfnamefont {W.-L.}\ \bibnamefont {Qian}},\ }\href {\doibase
  10.1103/PhysRevD.109.044031} {\bibfield  {journal} {\bibinfo  {journal}
  {Phys. Rev. D}\ }\textbf {\bibinfo {volume} {109}},\ \bibinfo {pages}
  {044031} (\bibinfo {year} {2024})},\ \Eprint
  {http://arxiv.org/abs/2308.15371} {arXiv:2308.15371 [gr-qc]} \BibitemShut
  {NoStop}%
\bibitem [{\citenamefont {Chen}\ and\ \citenamefont
  {Kotlařík}(2023)}]{Chen:2023akf}%
  \BibitemOpen
  \bibfield  {author} {\bibinfo {author} {\bibfnamefont {C.-Y.}\ \bibnamefont
  {Chen}}\ and\ \bibinfo {author} {\bibfnamefont {P.}~\bibnamefont
  {Kotlařík}},\ }\href {\doibase 10.1103/PhysRevD.108.064052} {\bibfield
  {journal} {\bibinfo  {journal} {Phys. Rev. D}\ }\textbf {\bibinfo {volume}
  {108}},\ \bibinfo {pages} {064052} (\bibinfo {year} {2023})},\ \Eprint
  {http://arxiv.org/abs/2307.07360} {arXiv:2307.07360 [gr-qc]} \BibitemShut
  {NoStop}%
\bibitem [{\citenamefont {Zhao}\ \emph {et~al.}(2023)\citenamefont {Zhao},
  \citenamefont {Sun}, \citenamefont {Lin},\ and\ \citenamefont
  {Cao}}]{Zhao:2023tyo}%
  \BibitemOpen
  \bibfield  {author} {\bibinfo {author} {\bibfnamefont {Y.}~\bibnamefont
  {Zhao}}, \bibinfo {author} {\bibfnamefont {B.}~\bibnamefont {Sun}}, \bibinfo
  {author} {\bibfnamefont {K.}~\bibnamefont {Lin}}, \ and\ \bibinfo {author}
  {\bibfnamefont {Z.}~\bibnamefont {Cao}},\ }\href {\doibase
  10.1103/PhysRevD.108.024070} {\bibfield  {journal} {\bibinfo  {journal}
  {Phys. Rev. D}\ }\textbf {\bibinfo {volume} {108}},\ \bibinfo {pages}
  {024070} (\bibinfo {year} {2023})},\ \Eprint
  {http://arxiv.org/abs/2303.09215} {arXiv:2303.09215 [gr-qc]} \BibitemShut
  {NoStop}%
\bibitem [{\citenamefont {Xavier}\ \emph {et~al.}(2023)\citenamefont {Xavier},
  \citenamefont {Lima},\ and\ \citenamefont {Crispino}}]{Xavier:2023exm}%
  \BibitemOpen
  \bibfield  {author} {\bibinfo {author} {\bibfnamefont {S.~V. M. C.~B.}\
  \bibnamefont {Xavier}}, \bibinfo {author} {\bibfnamefont {H.~C.~D.}\
  \bibnamefont {Lima}, \bibfnamefont {Junior.}}, \ and\ \bibinfo {author}
  {\bibfnamefont {L.~C.~B.}\ \bibnamefont {Crispino}},\ }\href {\doibase
  10.1103/PhysRevD.107.064040} {\bibfield  {journal} {\bibinfo  {journal}
  {Phys. Rev. D}\ }\textbf {\bibinfo {volume} {107}},\ \bibinfo {pages}
  {064040} (\bibinfo {year} {2023})},\ \Eprint
  {http://arxiv.org/abs/2303.17666} {arXiv:2303.17666 [gr-qc]} \BibitemShut
  {NoStop}%
\bibitem [{\citenamefont {Figueiredo}\ \emph {et~al.}(2023)\citenamefont
  {Figueiredo}, \citenamefont {Maselli},\ and\ \citenamefont
  {Cardoso}}]{Figueiredo:2023gas}%
  \BibitemOpen
  \bibfield  {author} {\bibinfo {author} {\bibfnamefont {E.}~\bibnamefont
  {Figueiredo}}, \bibinfo {author} {\bibfnamefont {A.}~\bibnamefont {Maselli}},
  \ and\ \bibinfo {author} {\bibfnamefont {V.}~\bibnamefont {Cardoso}},\ }\href
  {\doibase 10.1103/PhysRevD.107.104033} {\bibfield  {journal} {\bibinfo
  {journal} {Phys. Rev. D}\ }\textbf {\bibinfo {volume} {107}},\ \bibinfo
  {pages} {104033} (\bibinfo {year} {2023})},\ \Eprint
  {http://arxiv.org/abs/2303.08183} {arXiv:2303.08183 [gr-qc]} \BibitemShut
  {NoStop}%
\bibitem [{\citenamefont {Daghigh}\ and\ \citenamefont
  {Kunstatter}(2022)}]{Daghigh:2022pcr}%
  \BibitemOpen
  \bibfield  {author} {\bibinfo {author} {\bibfnamefont {R.~G.}\ \bibnamefont
  {Daghigh}}\ and\ \bibinfo {author} {\bibfnamefont {G.}~\bibnamefont
  {Kunstatter}},\ }\href {\doibase 10.3847/1538-4357/ac940b} {\bibfield
  {journal} {\bibinfo  {journal} {Astrophys. J.}\ }\textbf {\bibinfo {volume}
  {940}},\ \bibinfo {pages} {33} (\bibinfo {year} {2022})},\ \bibinfo {note}
  {[Erratum: Astrophys.J. 963, 167 (2024)]},\ \Eprint
  {http://arxiv.org/abs/2206.04195} {arXiv:2206.04195 [astro-ph.GA]}
  \BibitemShut {NoStop}%
\bibitem [{\citenamefont {Liu}\ \emph {et~al.}(2023)\citenamefont {Liu},
  \citenamefont {Yang}, \citenamefont {Övgün}, \citenamefont {Long},\ and\
  \citenamefont {Xu}}]{Liu:2022ygf}%
  \BibitemOpen
  \bibfield  {author} {\bibinfo {author} {\bibfnamefont {D.}~\bibnamefont
  {Liu}}, \bibinfo {author} {\bibfnamefont {Y.}~\bibnamefont {Yang}}, \bibinfo
  {author} {\bibfnamefont {A.}~\bibnamefont {Övgün}}, \bibinfo {author}
  {\bibfnamefont {Z.-W.}\ \bibnamefont {Long}}, \ and\ \bibinfo {author}
  {\bibfnamefont {Z.}~\bibnamefont {Xu}},\ }\href {\doibase
  10.1140/epjc/s10052-023-11739-w} {\bibfield  {journal} {\bibinfo  {journal}
  {Eur. Phys. J. C}\ }\textbf {\bibinfo {volume} {83}},\ \bibinfo {pages} {565}
  (\bibinfo {year} {2023})},\ \Eprint {http://arxiv.org/abs/2204.11563}
  {arXiv:2204.11563 [gr-qc]} \BibitemShut {NoStop}%
\bibitem [{\citenamefont {Liu}\ \emph {et~al.}(2022)\citenamefont {Liu},
  \citenamefont {Chen},\ and\ \citenamefont {Jing}}]{Liu:2022lrg}%
  \BibitemOpen
  \bibfield  {author} {\bibinfo {author} {\bibfnamefont {J.}~\bibnamefont
  {Liu}}, \bibinfo {author} {\bibfnamefont {S.}~\bibnamefont {Chen}}, \ and\
  \bibinfo {author} {\bibfnamefont {J.}~\bibnamefont {Jing}},\ }\href {\doibase
  10.1088/1674-1137/ac7856} {\bibfield  {journal} {\bibinfo  {journal} {Chin.
  Phys. C}\ }\textbf {\bibinfo {volume} {46}},\ \bibinfo {pages} {105104}
  (\bibinfo {year} {2022})},\ \Eprint {http://arxiv.org/abs/2203.14039}
  {arXiv:2203.14039 [gr-qc]} \BibitemShut {NoStop}%
\bibitem [{\citenamefont {Jusufi}(2023)}]{Jusufi:2022jxu}%
  \BibitemOpen
  \bibfield  {author} {\bibinfo {author} {\bibfnamefont {K.}~\bibnamefont
  {Jusufi}},\ }\href {\doibase 10.1140/epjc/s10052-023-11264-w} {\bibfield
  {journal} {\bibinfo  {journal} {Eur. Phys. J. C}\ }\textbf {\bibinfo {volume}
  {83}},\ \bibinfo {pages} {103} (\bibinfo {year} {2023})},\ \Eprint
  {http://arxiv.org/abs/2202.00010} {arXiv:2202.00010 [gr-qc]} \BibitemShut
  {NoStop}%
\bibitem [{\citenamefont {Igata}\ \emph {et~al.}(2023)\citenamefont {Igata},
  \citenamefont {Harada}, \citenamefont {Saida},\ and\ \citenamefont
  {Takamori}}]{Igata:2022rcm}%
  \BibitemOpen
  \bibfield  {author} {\bibinfo {author} {\bibfnamefont {T.}~\bibnamefont
  {Igata}}, \bibinfo {author} {\bibfnamefont {T.}~\bibnamefont {Harada}},
  \bibinfo {author} {\bibfnamefont {H.}~\bibnamefont {Saida}}, \ and\ \bibinfo
  {author} {\bibfnamefont {Y.}~\bibnamefont {Takamori}},\ }\href {\doibase
  10.1142/S0218271823501055} {\bibfield  {journal} {\bibinfo  {journal} {Int.
  J. Mod. Phys. D}\ }\textbf {\bibinfo {volume} {32}},\ \bibinfo {pages}
  {2350105} (\bibinfo {year} {2023})},\ \Eprint
  {http://arxiv.org/abs/2202.00202} {arXiv:2202.00202 [gr-qc]} \BibitemShut
  {NoStop}%
\bibitem [{\citenamefont {Stuchlík}\ and\ \citenamefont
  {Vrba}(2021)}]{Stuchlik:2021gwg}%
  \BibitemOpen
  \bibfield  {author} {\bibinfo {author} {\bibfnamefont {Z.}~\bibnamefont
  {Stuchlík}}\ and\ \bibinfo {author} {\bibfnamefont {J.}~\bibnamefont
  {Vrba}},\ }\href {\doibase 10.1088/1475-7516/2021/11/059} {\bibfield
  {journal} {\bibinfo  {journal} {JCAP}\ }\textbf {\bibinfo {volume} {11}},\
  \bibinfo {pages} {059} (\bibinfo {year} {2021})},\ \Eprint
  {http://arxiv.org/abs/2110.07411} {arXiv:2110.07411 [gr-qc]} \BibitemShut
  {NoStop}%
\bibitem [{\citenamefont {Zhang}\ \emph {et~al.}(2021)\citenamefont {Zhang},
  \citenamefont {Zhu},\ and\ \citenamefont {Wang}}]{Zhang:2021bdr}%
  \BibitemOpen
  \bibfield  {author} {\bibinfo {author} {\bibfnamefont {C.}~\bibnamefont
  {Zhang}}, \bibinfo {author} {\bibfnamefont {T.}~\bibnamefont {Zhu}}, \ and\
  \bibinfo {author} {\bibfnamefont {A.}~\bibnamefont {Wang}},\ }\href {\doibase
  10.1103/PhysRevD.104.124082} {\bibfield  {journal} {\bibinfo  {journal}
  {Phys. Rev. D}\ }\textbf {\bibinfo {volume} {104}},\ \bibinfo {pages}
  {124082} (\bibinfo {year} {2021})},\ \Eprint
  {http://arxiv.org/abs/2111.04966} {arXiv:2111.04966 [gr-qc]} \BibitemShut
  {NoStop}%
\bibitem [{\citenamefont {Konoplya}(2019)}]{Konoplya:2019sns}%
  \BibitemOpen
  \bibfield  {author} {\bibinfo {author} {\bibfnamefont {R.~A.}\ \bibnamefont
  {Konoplya}},\ }\href {\doibase 10.1016/j.physletb.2019.05.043} {\bibfield
  {journal} {\bibinfo  {journal} {Phys. Lett. B}\ }\textbf {\bibinfo {volume}
  {795}},\ \bibinfo {pages} {1} (\bibinfo {year} {2019})},\ \Eprint
  {http://arxiv.org/abs/1905.00064} {arXiv:1905.00064 [gr-qc]} \BibitemShut
  {NoStop}%
\bibitem [{\citenamefont {Jha}(2025{\natexlab{a}})}]{Jha:2025uie}%
  \BibitemOpen
  \bibfield  {author} {\bibinfo {author} {\bibfnamefont {S.~K.}\ \bibnamefont
  {Jha}},\ }\href {\doibase 10.1088/1475-7516/2025/09/069} {\bibfield
  {journal} {\bibinfo  {journal} {JCAP}\ }\textbf {\bibinfo {volume} {09}},\
  \bibinfo {pages} {069} (\bibinfo {year} {2025}{\natexlab{a}})},\ \Eprint
  {http://arxiv.org/abs/2506.21911} {arXiv:2506.21911 [gr-qc]} \BibitemShut
  {NoStop}%
\bibitem [{\citenamefont {Santos}\ \emph {et~al.}(2025)\citenamefont {Santos},
  \citenamefont {Vieira}, \citenamefont {da~Silva},\ and\ \citenamefont
  {Bezerra}}]{Santos:2025sun}%
  \BibitemOpen
  \bibfield  {author} {\bibinfo {author} {\bibfnamefont {L.~C.~N.}\
  \bibnamefont {Santos}}, \bibinfo {author} {\bibfnamefont {H.~S.}\
  \bibnamefont {Vieira}}, \bibinfo {author} {\bibfnamefont {F.~M.}\
  \bibnamefont {da~Silva}}, \ and\ \bibinfo {author} {\bibfnamefont {V.~B.}\
  \bibnamefont {Bezerra}},\ }\href {\doibase 10.1140/epjc/s10052-025-14789-4}
  {\bibfield  {journal} {\bibinfo  {journal} {Eur. Phys. J. C}\ }\textbf
  {\bibinfo {volume} {85}},\ \bibinfo {pages} {1036} (\bibinfo {year}
  {2025})},\ \Eprint {http://arxiv.org/abs/2506.21639} {arXiv:2506.21639
  [gr-qc]} \BibitemShut {NoStop}%
\bibitem [{\citenamefont {Ahmed}\ \emph {et~al.}(2025)\citenamefont {Ahmed},
  \citenamefont {Al-Badawi},\ and\ \citenamefont {Sakallı}}]{Ahmed:2025ttq}%
  \BibitemOpen
  \bibfield  {author} {\bibinfo {author} {\bibfnamefont {F.}~\bibnamefont
  {Ahmed}}, \bibinfo {author} {\bibfnamefont {A.}~\bibnamefont {Al-Badawi}}, \
  and\ \bibinfo {author} {\bibfnamefont {{\.I}.}~\bibnamefont {Sakallı}},\
  }\href {\doibase 10.1140/epjc/s10052-025-14723-8} {\bibfield  {journal}
  {\bibinfo  {journal} {Eur. Phys. J. C}\ }\textbf {\bibinfo {volume} {85}},\
  \bibinfo {pages} {984} (\bibinfo {year} {2025})},\ \Eprint
  {http://arxiv.org/abs/2506.18457} {arXiv:2506.18457 [gr-qc]} \BibitemShut
  {NoStop}%
\bibitem [{\citenamefont {Luo}\ \emph {et~al.}(2025)\citenamefont {Luo},
  \citenamefont {Tang},\ and\ \citenamefont {Xu}}]{Luo:2025xjb}%
  \BibitemOpen
  \bibfield  {author} {\bibinfo {author} {\bibfnamefont {Z.}~\bibnamefont
  {Luo}}, \bibinfo {author} {\bibfnamefont {M.}~\bibnamefont {Tang}}, \ and\
  \bibinfo {author} {\bibfnamefont {Z.}~\bibnamefont {Xu}},\ }\href@noop {} {\
  (\bibinfo {year} {2025})},\ \Eprint {http://arxiv.org/abs/2505.20115}
  {arXiv:2505.20115 [gr-qc]} \BibitemShut {NoStop}%
\bibitem [{\citenamefont {Jha}(2025{\natexlab{b}})}]{Jha:2025xjf}%
  \BibitemOpen
  \bibfield  {author} {\bibinfo {author} {\bibfnamefont {S.~K.}\ \bibnamefont
  {Jha}},\ }\href {\doibase 10.1088/1475-7516/2025/06/033} {\bibfield
  {journal} {\bibinfo  {journal} {JCAP}\ }\textbf {\bibinfo {volume} {06}},\
  \bibinfo {pages} {033} (\bibinfo {year} {2025}{\natexlab{b}})},\ \Eprint
  {http://arxiv.org/abs/2503.19938} {arXiv:2503.19938 [gr-qc]} \BibitemShut
  {NoStop}%
\bibitem [{\citenamefont {Haroon}\ and\ \citenamefont
  {Zhu}(2025)}]{Haroon:2025rzx}%
  \BibitemOpen
  \bibfield  {author} {\bibinfo {author} {\bibfnamefont {S.}~\bibnamefont
  {Haroon}}\ and\ \bibinfo {author} {\bibfnamefont {T.}~\bibnamefont {Zhu}},\
  }\href {\doibase 10.1103/ckdt-wtsl} {\bibfield  {journal} {\bibinfo
  {journal} {Phys. Rev. D}\ }\textbf {\bibinfo {volume} {112}},\ \bibinfo
  {pages} {044046} (\bibinfo {year} {2025})},\ \Eprint
  {http://arxiv.org/abs/2502.09171} {arXiv:2502.09171 [gr-qc]} \BibitemShut
  {NoStop}%
\bibitem [{\citenamefont {Konoplya}\ \emph {et~al.}(2025)\citenamefont
  {Konoplya}, \citenamefont {Khrabustovskyi}, \citenamefont {Kříž},\ and\
  \citenamefont {Zhidenko}}]{Konoplya:2025mvj}%
  \BibitemOpen
  \bibfield  {author} {\bibinfo {author} {\bibfnamefont {R.~A.}\ \bibnamefont
  {Konoplya}}, \bibinfo {author} {\bibfnamefont {A.}~\bibnamefont
  {Khrabustovskyi}}, \bibinfo {author} {\bibfnamefont {J.}~\bibnamefont
  {Kříž}}, \ and\ \bibinfo {author} {\bibfnamefont {A.}~\bibnamefont
  {Zhidenko}},\ }\href {\doibase 10.1088/1475-7516/2025/04/062} {\bibfield
  {journal} {\bibinfo  {journal} {JCAP}\ }\textbf {\bibinfo {volume} {04}},\
  \bibinfo {pages} {062} (\bibinfo {year} {2025})},\ \Eprint
  {http://arxiv.org/abs/2501.16134} {arXiv:2501.16134 [gr-qc]} \BibitemShut
  {NoStop}%
\bibitem [{\citenamefont {He}\ \emph {et~al.}(2025{\natexlab{a}})\citenamefont
  {He}, \citenamefont {Yang},\ and\ \citenamefont {Zeng}}]{He:2025rjq}%
  \BibitemOpen
  \bibfield  {author} {\bibinfo {author} {\bibfnamefont {K.-J.}\ \bibnamefont
  {He}}, \bibinfo {author} {\bibfnamefont {C.-Y.}\ \bibnamefont {Yang}}, \ and\
  \bibinfo {author} {\bibfnamefont {X.-X.}\ \bibnamefont {Zeng}},\ }\href@noop
  {} {\  (\bibinfo {year} {2025}{\natexlab{a}})},\ \Eprint
  {http://arxiv.org/abs/2501.06778} {arXiv:2501.06778 [astro-ph.HE]}
  \BibitemShut {NoStop}%
\bibitem [{\citenamefont {He}\ \emph {et~al.}(2025{\natexlab{b}})\citenamefont
  {He}, \citenamefont {Li}, \citenamefont {Yang},\ and\ \citenamefont
  {Zeng}}]{He:2024amh}%
  \BibitemOpen
  \bibfield  {author} {\bibinfo {author} {\bibfnamefont {K.-J.}\ \bibnamefont
  {He}}, \bibinfo {author} {\bibfnamefont {G.-P.}\ \bibnamefont {Li}}, \bibinfo
  {author} {\bibfnamefont {C.-Y.}\ \bibnamefont {Yang}}, \ and\ \bibinfo
  {author} {\bibfnamefont {X.-X.}\ \bibnamefont {Zeng}},\ }\href {\doibase
  10.1140/epjc/s10052-025-14391-8} {\bibfield  {journal} {\bibinfo  {journal}
  {Eur. Phys. J. C}\ }\textbf {\bibinfo {volume} {85}},\ \bibinfo {pages} {662}
  (\bibinfo {year} {2025}{\natexlab{b}})},\ \Eprint
  {http://arxiv.org/abs/2411.11680} {arXiv:2411.11680 [astro-ph.HE]}
  \BibitemShut {NoStop}%
\bibitem [{\citenamefont {Zare}\ \emph {et~al.}(2024)\citenamefont {Zare},
  \citenamefont {Nieto}, \citenamefont {Feng}, \citenamefont {Dong},\ and\
  \citenamefont {Hassanabadi}}]{Zare:2024dtf}%
  \BibitemOpen
  \bibfield  {author} {\bibinfo {author} {\bibfnamefont {S.}~\bibnamefont
  {Zare}}, \bibinfo {author} {\bibfnamefont {L.~M.}\ \bibnamefont {Nieto}},
  \bibinfo {author} {\bibfnamefont {X.-H.}\ \bibnamefont {Feng}}, \bibinfo
  {author} {\bibfnamefont {S.-H.}\ \bibnamefont {Dong}}, \ and\ \bibinfo
  {author} {\bibfnamefont {H.}~\bibnamefont {Hassanabadi}},\ }\href {\doibase
  10.1088/1475-7516/2024/08/041} {\bibfield  {journal} {\bibinfo  {journal}
  {JCAP}\ }\textbf {\bibinfo {volume} {08}},\ \bibinfo {pages} {041} (\bibinfo
  {year} {2024})},\ \Eprint {http://arxiv.org/abs/2406.07300} {arXiv:2406.07300
  [astro-ph.HE]} \BibitemShut {NoStop}%
\bibitem [{\citenamefont {Bécar}\ \emph {et~al.}(2024)\citenamefont {Bécar},
  \citenamefont {González}, \citenamefont {Papantonopoulos},\ and\
  \citenamefont {Vásquez}}]{Becar:2023zbl}%
  \BibitemOpen
  \bibfield  {author} {\bibinfo {author} {\bibfnamefont {R.}~\bibnamefont
  {Bécar}}, \bibinfo {author} {\bibfnamefont {P.~A.}\ \bibnamefont
  {González}}, \bibinfo {author} {\bibfnamefont {E.}~\bibnamefont
  {Papantonopoulos}}, \ and\ \bibinfo {author} {\bibfnamefont {Y.}~\bibnamefont
  {Vásquez}},\ }\href {\doibase 10.1140/epjc/s10052-024-12553-8} {\bibfield
  {journal} {\bibinfo  {journal} {Eur. Phys. J. C}\ }\textbf {\bibinfo {volume}
  {84}},\ \bibinfo {pages} {329} (\bibinfo {year} {2024})},\ \Eprint
  {http://arxiv.org/abs/2310.00857} {arXiv:2310.00857 [gr-qc]} \BibitemShut
  {NoStop}%
\bibitem [{\citenamefont {Pantig}\ and\ \citenamefont
  {Övgün}(2023)}]{Pantig:2022sjb}%
  \BibitemOpen
  \bibfield  {author} {\bibinfo {author} {\bibfnamefont {R.~C.}\ \bibnamefont
  {Pantig}}\ and\ \bibinfo {author} {\bibfnamefont {A.}~\bibnamefont
  {Övgün}},\ }\href {\doibase 10.1002/prop.202200164} {\bibfield  {journal}
  {\bibinfo  {journal} {Fortsch. Phys.}\ }\textbf {\bibinfo {volume} {71}},\
  \bibinfo {pages} {2200164} (\bibinfo {year} {2023})},\ \Eprint
  {http://arxiv.org/abs/2210.00523} {arXiv:2210.00523 [gr-qc]} \BibitemShut
  {NoStop}%
\bibitem [{\citenamefont {Stuchlík}\ and\ \citenamefont
  {Vrba}(2022)}]{Stuchlik:2022xtq}%
  \BibitemOpen
  \bibfield  {author} {\bibinfo {author} {\bibfnamefont {Z.}~\bibnamefont
  {Stuchlík}}\ and\ \bibinfo {author} {\bibfnamefont {J.}~\bibnamefont
  {Vrba}},\ }\href {\doibase 10.3847/1538-4357/ac7f27} {\bibfield  {journal}
  {\bibinfo  {journal} {Astrophys. J.}\ }\textbf {\bibinfo {volume} {935}},\
  \bibinfo {pages} {91} (\bibinfo {year} {2022})},\ \Eprint
  {http://arxiv.org/abs/2208.02612} {arXiv:2208.02612 [gr-qc]} \BibitemShut
  {NoStop}%
\bibitem [{\citenamefont {Ravanal}\ \emph {et~al.}(2024)\citenamefont
  {Ravanal}, \citenamefont {Gómez},\ and\ \citenamefont
  {Cruz}}]{Ravanal:2024odh}%
  \BibitemOpen
  \bibfield  {author} {\bibinfo {author} {\bibfnamefont {Y.}~\bibnamefont
  {Ravanal}}, \bibinfo {author} {\bibfnamefont {G.}~\bibnamefont {Gómez}}, \
  and\ \bibinfo {author} {\bibfnamefont {N.}~\bibnamefont {Cruz}},\ }\href
  {\doibase 10.1103/PhysRevD.110.023027} {\bibfield  {journal} {\bibinfo
  {journal} {Phys. Rev. D}\ }\textbf {\bibinfo {volume} {110}},\ \bibinfo
  {pages} {023027} (\bibinfo {year} {2024})},\ \Eprint
  {http://arxiv.org/abs/2404.06774} {arXiv:2404.06774 [astro-ph.CO]}
  \BibitemShut {NoStop}%
\bibitem [{\citenamefont {Pantig}(2024)}]{Pantig:2024rmr}%
  \BibitemOpen
  \bibfield  {author} {\bibinfo {author} {\bibfnamefont {R.~C.}\ \bibnamefont
  {Pantig}},\ }\href {\doibase 10.1016/j.dark.2024.101550} {\bibfield
  {journal} {\bibinfo  {journal} {Phys. Dark Univ.}\ }\textbf {\bibinfo
  {volume} {45}},\ \bibinfo {pages} {101550} (\bibinfo {year} {2024})},\
  \Eprint {http://arxiv.org/abs/2405.07531} {arXiv:2405.07531 [gr-qc]}
  \BibitemShut {NoStop}%
\bibitem [{\citenamefont {Övgün}\ and\ \citenamefont
  {Pantig}(2025)}]{Ovgun:2025bol}%
  \BibitemOpen
  \bibfield  {author} {\bibinfo {author} {\bibfnamefont {A.}~\bibnamefont
  {Övgün}}\ and\ \bibinfo {author} {\bibfnamefont {R.~C.}\ \bibnamefont
  {Pantig}},\ }\href {\doibase 10.1016/j.physletb.2025.139398} {\bibfield
  {journal} {\bibinfo  {journal} {Phys. Lett. B}\ }\textbf {\bibinfo {volume}
  {864}},\ \bibinfo {pages} {139398} (\bibinfo {year} {2025})},\ \Eprint
  {http://arxiv.org/abs/2501.12559} {arXiv:2501.12559 [gr-qc]} \BibitemShut
  {NoStop}%
\bibitem [{\citenamefont {Bekenstein}(1971)}]{Bekenstein:1971ej}%
  \BibitemOpen
  \bibfield  {author} {\bibinfo {author} {\bibfnamefont {J.~D.}\ \bibnamefont
  {Bekenstein}},\ }\href {\doibase 10.1103/PhysRevD.4.2185} {\bibfield
  {journal} {\bibinfo  {journal} {Phys. Rev. D}\ }\textbf {\bibinfo {volume}
  {4}},\ \bibinfo {pages} {2185} (\bibinfo {year} {1971})}\BibitemShut
  {NoStop}%
\bibitem [{\citenamefont {Wald}(1974)}]{Wald:1974np}%
  \BibitemOpen
  \bibfield  {author} {\bibinfo {author} {\bibfnamefont {R.~M.}\ \bibnamefont
  {Wald}},\ }\href {\doibase 10.1103/PhysRevD.10.1680} {\bibfield  {journal}
  {\bibinfo  {journal} {Phys. Rev. D}\ }\textbf {\bibinfo {volume} {10}},\
  \bibinfo {pages} {1680} (\bibinfo {year} {1974})}\BibitemShut {NoStop}%
\bibitem [{\citenamefont {Ruffini}\ and\ \citenamefont
  {Wilson}(1975)}]{Ruffini:1975ne}%
  \BibitemOpen
  \bibfield  {author} {\bibinfo {author} {\bibfnamefont {R.}~\bibnamefont
  {Ruffini}}\ and\ \bibinfo {author} {\bibfnamefont {J.~R.}\ \bibnamefont
  {Wilson}},\ }\href {\doibase 10.1103/PhysRevD.12.2959} {\bibfield  {journal}
  {\bibinfo  {journal} {Phys. Rev. D}\ }\textbf {\bibinfo {volume} {12}},\
  \bibinfo {pages} {2959} (\bibinfo {year} {1975})}\BibitemShut {NoStop}%
\bibitem [{\citenamefont {Tursunov}\ \emph {et~al.}(2020)\citenamefont
  {Tursunov}, \citenamefont {Zajaček}, \citenamefont {Eckart},\ and\
  \citenamefont {Stuchlík}}]{Tursunov:2019mox}%
  \BibitemOpen
  \bibfield  {author} {\bibinfo {author} {\bibfnamefont {A.}~\bibnamefont
  {Tursunov}}, \bibinfo {author} {\bibfnamefont {M.}~\bibnamefont {Zajaček}},
  \bibinfo {author} {\bibfnamefont {A.}~\bibnamefont {Eckart}}, \ and\ \bibinfo
  {author} {\bibfnamefont {Z.}~\bibnamefont {Stuchlík}},\ }\href {\doibase
  10.3847/1538-4357/ab980e} {\bibfield  {journal} {\bibinfo  {journal}
  {Astrophys. J.}\ }\textbf {\bibinfo {volume} {897}},\ \bibinfo {pages} {99}
  (\bibinfo {year} {2020})},\ \Eprint {http://arxiv.org/abs/1912.08174}
  {arXiv:1912.08174 [astro-ph.GA]} \BibitemShut {NoStop}%
\bibitem [{\citenamefont {Juraev}\ \emph {et~al.}(2024)\citenamefont {Juraev},
  \citenamefont {Stuchlík}, \citenamefont {Tursunov},\ and\ \citenamefont
  {Kološ}}]{Juraev:2024dju}%
  \BibitemOpen
  \bibfield  {author} {\bibinfo {author} {\bibfnamefont {B.}~\bibnamefont
  {Juraev}}, \bibinfo {author} {\bibfnamefont {Z.}~\bibnamefont {Stuchlík}},
  \bibinfo {author} {\bibfnamefont {A.}~\bibnamefont {Tursunov}}, \ and\
  \bibinfo {author} {\bibfnamefont {M.}~\bibnamefont {Kološ}},\ }\href
  {\doibase 10.1088/1475-7516/2024/09/035} {\bibfield  {journal} {\bibinfo
  {journal} {JCAP}\ }\textbf {\bibinfo {volume} {09}},\ \bibinfo {pages} {035}
  (\bibinfo {year} {2024})},\ \Eprint {http://arxiv.org/abs/2402.13797}
  {arXiv:2402.13797 [gr-qc]} \BibitemShut {NoStop}%
\bibitem [{\citenamefont {Hernquist}(1990)}]{Hernquist:1990be}%
  \BibitemOpen
  \bibfield  {author} {\bibinfo {author} {\bibfnamefont {L.}~\bibnamefont
  {Hernquist}},\ }\href {\doibase 10.1086/168845} {\bibfield  {journal}
  {\bibinfo  {journal} {Astrophys. J.}\ }\textbf {\bibinfo {volume} {356}},\
  \bibinfo {pages} {359} (\bibinfo {year} {1990})}\BibitemShut {NoStop}%
\bibitem [{\citenamefont {Burkert}(1995)}]{Burkert:1995yz}%
  \BibitemOpen
  \bibfield  {author} {\bibinfo {author} {\bibfnamefont {A.}~\bibnamefont
  {Burkert}},\ }\href {\doibase 10.1086/309560} {\bibfield  {journal} {\bibinfo
   {journal} {Astrophys. J. Lett.}\ }\textbf {\bibinfo {volume} {447}},\
  \bibinfo {pages} {L25} (\bibinfo {year} {1995})},\ \Eprint
  {http://arxiv.org/abs/astro-ph/9504041} {arXiv:astro-ph/9504041} \BibitemShut
  {NoStop}%
\bibitem [{\citenamefont {Navarro}\ \emph {et~al.}(1995)\citenamefont
  {Navarro}, \citenamefont {Frenk},\ and\ \citenamefont
  {White}}]{Navarro:1994hi}%
  \BibitemOpen
  \bibfield  {author} {\bibinfo {author} {\bibfnamefont {J.~F.}\ \bibnamefont
  {Navarro}}, \bibinfo {author} {\bibfnamefont {C.~S.}\ \bibnamefont {Frenk}},
  \ and\ \bibinfo {author} {\bibfnamefont {S.~D.~M.}\ \bibnamefont {White}},\
  }\href {\doibase 10.1093/mnras/275.3.720} {\bibfield  {journal} {\bibinfo
  {journal} {Mon. Not. Roy. Astron. Soc.}\ }\textbf {\bibinfo {volume} {275}},\
  \bibinfo {pages} {720} (\bibinfo {year} {1995})},\ \Eprint
  {http://arxiv.org/abs/astro-ph/9408069} {arXiv:astro-ph/9408069} \BibitemShut
  {NoStop}%
\bibitem [{\citenamefont {Taylor}\ and\ \citenamefont
  {Silk}(2003)}]{Taylor:2002zd}%
  \BibitemOpen
  \bibfield  {author} {\bibinfo {author} {\bibfnamefont {J.~E.}\ \bibnamefont
  {Taylor}}\ and\ \bibinfo {author} {\bibfnamefont {J.}~\bibnamefont {Silk}},\
  }\href {\doibase 10.1046/j.1365-8711.2003.06201.x} {\bibfield  {journal}
  {\bibinfo  {journal} {Mon. Not. Roy. Astron. Soc.}\ }\textbf {\bibinfo
  {volume} {339}},\ \bibinfo {pages} {505} (\bibinfo {year} {2003})},\ \Eprint
  {http://arxiv.org/abs/astro-ph/0207299} {arXiv:astro-ph/0207299} \BibitemShut
  {NoStop}%
\bibitem [{\citenamefont {Moore}\ \emph {et~al.}(1998)\citenamefont {Moore},
  \citenamefont {Governato}, \citenamefont {Quinn}, \citenamefont {Stadel},\
  and\ \citenamefont {Lake}}]{Moore:1997sg}%
  \BibitemOpen
  \bibfield  {author} {\bibinfo {author} {\bibfnamefont {B.}~\bibnamefont
  {Moore}}, \bibinfo {author} {\bibfnamefont {F.}~\bibnamefont {Governato}},
  \bibinfo {author} {\bibfnamefont {T.~R.}\ \bibnamefont {Quinn}}, \bibinfo
  {author} {\bibfnamefont {J.}~\bibnamefont {Stadel}}, \ and\ \bibinfo {author}
  {\bibfnamefont {G.}~\bibnamefont {Lake}},\ }\href {\doibase 10.1086/311333}
  {\bibfield  {journal} {\bibinfo  {journal} {Astrophys. J. Lett.}\ }\textbf
  {\bibinfo {volume} {499}},\ \bibinfo {pages} {L5} (\bibinfo {year} {1998})},\
  \Eprint {http://arxiv.org/abs/astro-ph/9709051} {arXiv:astro-ph/9709051}
  \BibitemShut {NoStop}%
\bibitem [{\citenamefont {Konoplya}\ and\ \citenamefont
  {Zhidenko}(2022{\natexlab{a}})}]{Konoplya:2022hbl}%
  \BibitemOpen
  \bibfield  {author} {\bibinfo {author} {\bibfnamefont {R.~A.}\ \bibnamefont
  {Konoplya}}\ and\ \bibinfo {author} {\bibfnamefont {A.}~\bibnamefont
  {Zhidenko}},\ }\href {\doibase 10.3847/1538-4357/ac76bc} {\bibfield
  {journal} {\bibinfo  {journal} {Astrophys. J.}\ }\textbf {\bibinfo {volume}
  {933}},\ \bibinfo {pages} {166} (\bibinfo {year} {2022}{\natexlab{a}})},\
  \Eprint {http://arxiv.org/abs/2202.02205} {arXiv:2202.02205 [gr-qc]}
  \BibitemShut {NoStop}%
\bibitem [{\citenamefont {Benson}(2010)}]{Benson:2010de}%
  \BibitemOpen
  \bibfield  {author} {\bibinfo {author} {\bibfnamefont {A.~J.}\ \bibnamefont
  {Benson}},\ }\href {\doibase 10.1016/j.physrep.2010.06.001} {\bibfield
  {journal} {\bibinfo  {journal} {Phys. Rept.}\ }\textbf {\bibinfo {volume}
  {495}},\ \bibinfo {pages} {33} (\bibinfo {year} {2010})},\ \Eprint
  {http://arxiv.org/abs/1006.5394} {arXiv:1006.5394 [astro-ph.CO]} \BibitemShut
  {NoStop}%
\bibitem [{\citenamefont {Dehnen}(1993)}]{Dehnen:1993uh}%
  \BibitemOpen
  \bibfield  {author} {\bibinfo {author} {\bibfnamefont {W.}~\bibnamefont
  {Dehnen}},\ }\href@noop {} {\bibfield  {journal} {\bibinfo  {journal} {Mon.
  Not. Roy. Astron. Soc.}\ }\textbf {\bibinfo {volume} {265}},\ \bibinfo
  {pages} {250} (\bibinfo {year} {1993})}\BibitemShut {NoStop}%
\bibitem [{\citenamefont {Bertone}\ \emph {et~al.}(2005)\citenamefont
  {Bertone}, \citenamefont {Hooper},\ and\ \citenamefont
  {Silk}}]{Bertone:2004pz}%
  \BibitemOpen
  \bibfield  {author} {\bibinfo {author} {\bibfnamefont {G.}~\bibnamefont
  {Bertone}}, \bibinfo {author} {\bibfnamefont {D.}~\bibnamefont {Hooper}}, \
  and\ \bibinfo {author} {\bibfnamefont {J.}~\bibnamefont {Silk}},\ }\href
  {\doibase 10.1016/j.physrep.2004.08.031} {\bibfield  {journal} {\bibinfo
  {journal} {Phys. Rept.}\ }\textbf {\bibinfo {volume} {405}},\ \bibinfo
  {pages} {279} (\bibinfo {year} {2005})},\ \Eprint
  {http://arxiv.org/abs/hep-ph/0404175} {hep-ph/0404175} \BibitemShut {NoStop}%
\bibitem [{\citenamefont {Tumlinson}\ \emph {et~al.}(2017)\citenamefont
  {Tumlinson}, \citenamefont {Peeples},\ and\ \citenamefont
  {Werk}}]{Tumlinson_2017}%
  \BibitemOpen
  \bibfield  {author} {\bibinfo {author} {\bibfnamefont {J.}~\bibnamefont
  {Tumlinson}}, \bibinfo {author} {\bibfnamefont {M.~S.}\ \bibnamefont
  {Peeples}}, \ and\ \bibinfo {author} {\bibfnamefont {J.~K.}\ \bibnamefont
  {Werk}},\ }\href {\doibase 10.1146/annurev-astro-091916-055240} {\bibfield
  {journal} {\bibinfo  {journal} {Annual Review of Astronomy and Astrophysics}\
  }\textbf {\bibinfo {volume} {55}},\ \bibinfo {pages} {389–432} (\bibinfo
  {year} {2017})}\BibitemShut {NoStop}%
\bibitem [{\citenamefont {Konoplya}\ and\ \citenamefont
  {Zhidenko}(2011)}]{Konoplya:2011qq}%
  \BibitemOpen
  \bibfield  {author} {\bibinfo {author} {\bibfnamefont {R.~A.}\ \bibnamefont
  {Konoplya}}\ and\ \bibinfo {author} {\bibfnamefont {A.}~\bibnamefont
  {Zhidenko}},\ }\href {\doibase 10.1103/RevModPhys.83.793} {\bibfield
  {journal} {\bibinfo  {journal} {Rev. Mod. Phys.}\ }\textbf {\bibinfo {volume}
  {83}},\ \bibinfo {pages} {793} (\bibinfo {year} {2011})},\ \Eprint
  {http://arxiv.org/abs/1102.4014} {arXiv:1102.4014 [gr-qc]} \BibitemShut
  {NoStop}%
\bibitem [{\citenamefont {Bolokhov}\ and\ \citenamefont
  {Skvortsova}(2025)}]{Bolokhov:2025uxz}%
  \BibitemOpen
  \bibfield  {author} {\bibinfo {author} {\bibfnamefont {S.~V.}\ \bibnamefont
  {Bolokhov}}\ and\ \bibinfo {author} {\bibfnamefont {M.}~\bibnamefont
  {Skvortsova}},\ }\href@noop {} {\  (\bibinfo {year} {2025})},\ \Eprint
  {http://arxiv.org/abs/2504.05014} {arXiv:2504.05014 [gr-qc]} \BibitemShut
  {NoStop}%
\bibitem [{\citenamefont {Cardoso}\ \emph {et~al.}(2009)\citenamefont
  {Cardoso}, \citenamefont {Miranda}, \citenamefont {Berti}, \citenamefont
  {Witek},\ and\ \citenamefont {Zanchin}}]{Cardoso:2008bp}%
  \BibitemOpen
  \bibfield  {author} {\bibinfo {author} {\bibfnamefont {V.}~\bibnamefont
  {Cardoso}}, \bibinfo {author} {\bibfnamefont {A.~S.}\ \bibnamefont
  {Miranda}}, \bibinfo {author} {\bibfnamefont {E.}~\bibnamefont {Berti}},
  \bibinfo {author} {\bibfnamefont {H.}~\bibnamefont {Witek}}, \ and\ \bibinfo
  {author} {\bibfnamefont {V.~T.}\ \bibnamefont {Zanchin}},\ }\href {\doibase
  10.1103/PhysRevD.79.064016} {\bibfield  {journal} {\bibinfo  {journal} {Phys.
  Rev. D}\ }\textbf {\bibinfo {volume} {79}},\ \bibinfo {pages} {064016}
  (\bibinfo {year} {2009})},\ \Eprint {http://arxiv.org/abs/0812.1806}
  {arXiv:0812.1806 [hep-th]} \BibitemShut {NoStop}%
\bibitem [{\citenamefont {Bolokhov}(2024)}]{Bolokhov:2023dxq}%
  \BibitemOpen
  \bibfield  {author} {\bibinfo {author} {\bibfnamefont {S.~V.}\ \bibnamefont
  {Bolokhov}},\ }\href {\doibase 10.1016/j.physletb.2024.138879} {\bibfield
  {journal} {\bibinfo  {journal} {Phys. Lett. B}\ }\textbf {\bibinfo {volume}
  {856}},\ \bibinfo {pages} {138879} (\bibinfo {year} {2024})},\ \Eprint
  {http://arxiv.org/abs/2310.12326} {arXiv:2310.12326 [gr-qc]} \BibitemShut
  {NoStop}%
\bibitem [{\citenamefont {Konoplya}(2023)}]{Konoplya:2022gjp}%
  \BibitemOpen
  \bibfield  {author} {\bibinfo {author} {\bibfnamefont {R.~A.}\ \bibnamefont
  {Konoplya}},\ }\href {\doibase 10.1016/j.physletb.2023.137674} {\bibfield
  {journal} {\bibinfo  {journal} {Phys. Lett. B}\ }\textbf {\bibinfo {volume}
  {838}},\ \bibinfo {pages} {137674} (\bibinfo {year} {2023})},\ \Eprint
  {http://arxiv.org/abs/2210.08373} {arXiv:2210.08373 [gr-qc]} \BibitemShut
  {NoStop}%
\bibitem [{\citenamefont {Konoplya}\ \emph {et~al.}(2019)\citenamefont
  {Konoplya}, \citenamefont {Zinhailo},\ and\ \citenamefont
  {Stuchlík}}]{Konoplya:2019hml}%
  \BibitemOpen
  \bibfield  {author} {\bibinfo {author} {\bibfnamefont {R.~A.}\ \bibnamefont
  {Konoplya}}, \bibinfo {author} {\bibfnamefont {A.~F.}\ \bibnamefont
  {Zinhailo}}, \ and\ \bibinfo {author} {\bibfnamefont {Z.}~\bibnamefont
  {Stuchlík}},\ }\href {\doibase 10.1103/PhysRevD.99.124042} {\bibfield
  {journal} {\bibinfo  {journal} {Phys. Rev. D}\ }\textbf {\bibinfo {volume}
  {99}},\ \bibinfo {pages} {124042} (\bibinfo {year} {2019})},\ \Eprint
  {http://arxiv.org/abs/1903.03483} {arXiv:1903.03483 [gr-qc]} \BibitemShut
  {NoStop}%
\bibitem [{\citenamefont {Konoplya}\ and\ \citenamefont
  {Zinhailo}(2020)}]{Konoplya:2020bxa}%
  \BibitemOpen
  \bibfield  {author} {\bibinfo {author} {\bibfnamefont {R.~A.}\ \bibnamefont
  {Konoplya}}\ and\ \bibinfo {author} {\bibfnamefont {A.~F.}\ \bibnamefont
  {Zinhailo}},\ }\href {\doibase 10.1140/epjc/s10052-020-08639-8} {\bibfield
  {journal} {\bibinfo  {journal} {Eur. Phys. J. C}\ }\textbf {\bibinfo {volume}
  {80}},\ \bibinfo {pages} {1049} (\bibinfo {year} {2020})},\ \Eprint
  {http://arxiv.org/abs/2003.01188} {arXiv:2003.01188 [gr-qc]} \BibitemShut
  {NoStop}%
\bibitem [{\citenamefont {Konoplya}\ and\ \citenamefont
  {Stuchlík}(2017)}]{Konoplya:2017wot}%
  \BibitemOpen
  \bibfield  {author} {\bibinfo {author} {\bibfnamefont {R.~A.}\ \bibnamefont
  {Konoplya}}\ and\ \bibinfo {author} {\bibfnamefont {Z.}~\bibnamefont
  {Stuchlík}},\ }\href {\doibase 10.1016/j.physletb.2017.06.015} {\bibfield
  {journal} {\bibinfo  {journal} {Phys. Lett. B}\ }\textbf {\bibinfo {volume}
  {771}},\ \bibinfo {pages} {597} (\bibinfo {year} {2017})},\ \Eprint
  {http://arxiv.org/abs/1705.05928} {arXiv:1705.05928 [gr-qc]} \BibitemShut
  {NoStop}%
\bibitem [{\citenamefont {Konoplya}\ and\ \citenamefont
  {Zhidenko}(2022{\natexlab{b}})}]{Konoplya:2022xid}%
  \BibitemOpen
  \bibfield  {author} {\bibinfo {author} {\bibfnamefont {R.~A.}\ \bibnamefont
  {Konoplya}}\ and\ \bibinfo {author} {\bibfnamefont {A.}~\bibnamefont
  {Zhidenko}},\ }\href {\doibase 10.1103/PhysRevD.106.124004} {\bibfield
  {journal} {\bibinfo  {journal} {Phys. Rev. D}\ }\textbf {\bibinfo {volume}
  {106}},\ \bibinfo {pages} {124004} (\bibinfo {year} {2022}{\natexlab{b}})},\
  \Eprint {http://arxiv.org/abs/2209.12058} {arXiv:2209.12058 [gr-qc]}
  \BibitemShut {NoStop}%
\bibitem [{\citenamefont {Konoplya}(2024)}]{Konoplya:2024ptj}%
  \BibitemOpen
  \bibfield  {author} {\bibinfo {author} {\bibfnamefont {R.~A.}\ \bibnamefont
  {Konoplya}},\ }\href {\doibase 10.1103/PhysRevD.109.104018} {\bibfield
  {journal} {\bibinfo  {journal} {Phys. Rev. D}\ }\textbf {\bibinfo {volume}
  {109}},\ \bibinfo {pages} {104018} (\bibinfo {year} {2024})},\ \Eprint
  {http://arxiv.org/abs/2401.17106} {arXiv:2401.17106 [gr-qc]} \BibitemShut
  {NoStop}%
\bibitem [{\citenamefont {Stuchlík}\ and\ \citenamefont
  {Zhidenko}(2025)}]{Stuchlik:2025mjj}%
  \BibitemOpen
  \bibfield  {author} {\bibinfo {author} {\bibfnamefont {Z.}~\bibnamefont
  {Stuchlík}}\ and\ \bibinfo {author} {\bibfnamefont {A.}~\bibnamefont
  {Zhidenko}},\ }\href@noop {} {\  (\bibinfo {year} {2025})},\ \Eprint
  {http://arxiv.org/abs/2506.09829} {arXiv:2506.09829 [gr-qc]} \BibitemShut
  {NoStop}%
\end{thebibliography}%

\end{document}